# Decoding Patterns of Data Generation Teams for Clinical and Scientific Success

Insights from the Bridge2AI Talent Knowledge Graph

Jiawei Xu
School of Information
University of Texas at Austin
Austin, Texas, USA
jiaweixu@utexas.edu

Qingnan Xie
Center for Labor and a Just Economy
Harvard Law School
Cambridge, Massachusetts, USA
qingnanx@gmail.com

Meijun Liu
Institute for Global Public Policy
Fudan University
Shanghai, China
meijunliu@fudan.edu.cn

Zhandos Sembay
School of Medicine
University of Alabama at Birmingham
Birmingham, Alabama, USA
zsembay8@uab.edu

Swathi Thaker
School of Medicine
University of Alabama at Birmingham
Birmingham, Alabama, USA
snthaker@uab.edu

Pamela Payne-Foster
School of Medicine
University of Alabama
Tuscaloosa, Alabama, USA
pfoster@ua.edu

Jake Chen
School of Medicine
University of Alabama at Birmingham
Birmingham, Alabama, USA
jakechen@uab.edu

Ying Ding
School of Information
University of Texas at Austin
Austin, TX, USA
ying.ding@ischool.utexas.edu

**ABSTRACT**

High-quality biomedical datasets are essential for medical research and disease treatment innovation. The NIH-funded Bridge2AI project strives to facilitate such innovations by uniting top-tier, diverse teams to curate datasets designed for AI-driven biomedical research. In this study, we examined 1,699 dataset papers from the Nucleic Acids Research (NAR) database issues and the Bridge2AI Talent Knowledge Graph. By treating each paper's authors as a team, we explored the relationship between team attributes (team power and fairness) and dataset paper quality, measured by scientific impact (Relative Citation Ratio percentile) and clinical translation power (APT, likelihood of citation by clinical trials and guidelines). Utilizing the SHAP explainable AI framework, we identified correlations between team attributes and the success of dataset papers in both citation impact and clinical translation. Key findings reveal that (1) PI (Principal Investigator) leadership and team academic prowess are strong predictors of dataset success; (2) team size and career age are positively correlated with scientific impact but show inverse patterns for clinical translation; and (3) higher female representation correlates with greater dataset success. Although our results are correlational, they offer valuable insights into forming high-performing data generation teams. Future research should incorporate causal frameworks to deepen understanding of these relationships.

## 1 Introduction

Biomedical research and health practices hinge on high-quality datasets—repositories of meticulously curated information that drive both scientific discovery and clinical innovation [1-3]. The integrity and representativeness of these datasets are essential, as they must serve as unbiased and credible foundations for advancements across the biomedical field. The Bridge2AI project [4], funded by the National Institutes of Health (NIH), leads the charge in this endeavor, focusing on creating and curating top-tier biomedical datasets. Researchers adhere to the FAIR Principles [5] to ensure databases are findable and reusable by machines. Each year, the Nucleic Acids Research (NAR) journal publishes a special issue detailing high-value online database resources for the biological community [6]. As the demand for AI-ready biomedical datasets grows, efforts to make these datasets ethically and reliably processable by AI are intensifying [7].

Studies show that team attributes, such as team leadership [8], [9], gender [10], and ethnicity diversity [11, 12], are significant to the quality of scientific findings and medical practices [13]. However, work is sparse on the relationship between team attributes and the scientific impact and translation potentials of biomedical datasets in subsequent medical research and practices. It is imperative to understand the patterns within data generation teams that contribute to the creation of high-quality, AI-ready datasets. We implemented the Relative Citation Ratio (RCR) [14] of papers that publish biomedical datasets to indicate the scientific impact of datasets. We implemented the Approximate Potential to Translate (APT) [15] as the measurement of the clinical translation power of dataset papers. In the biomedical field, the clinical translation of research outputs is critical while it often takes decades for medical advances to impact clinical practices. The APT metric, which predicts the potential knowledge flows from research papers to future clinical trials and guidelines, estimates the potential impact of a biomedical dataset on future medical practices [15].

We examined the dynamics of 1,699 biomedical dataset papers published in NAR database issues based on the Bridge2AI Talent Knowledge Graph (TKG) [16]. Our goal was to uncover the relationship between team attributes and dataset quality, assessed through scientific impact (RCR citation metrics) and clinical translational power (APT, predicted likelihood of citations by future clinical trials and guidelines). This study makes contributions by (1) identifying patterns of team power and fairness that correlate with the scientific impact and clinical translation and (2) enriching the Bridge2AI TKG with comprehensive team demographic information (gender, race) and clinical translational measurement (APT), making it suitable for researching team power and fairness in biomedical research.

## 2 Data and Method

### 2.1 Bridge2AI Talent Knowledge Graph

The Bridge2AI project [4] aims to create high-quality biomedical datasets optimized for machine learning to tackle complex biomedical challenges. This initiative unites experts from biomedicine, technology, and social sciences to produce FAIR and ethically sourced datasets. Bridge2AI TKG [16] offers an ideal context for examining teaming patterns—such as team leadership, structure, and diversity—and their relationships with scientific outcomes and clinical translation. With the 2024 version of PubMed Knowledge Graph [17], we extended the Bridge2AI TKG to include 147 Bridge2AI core researchers, their 44,000 co-authors, and their collective output of 2.05 million papers. The enriched Bridge2AI TKG contains bibliometric data, author information (including gender and race), scientific impact metrics, and clinical translation power metrics.

### 2.2 Distinguish Biomedical Dataset Papers

To identify specific data generation teams within the Bridge2AI TKG, we focused on publications in the NAR database issues, which publish high-value online database resources for the biological community [19]. Since 2004, NAR has tagged such issues with the identifier 'D1' or 'database issue'. Using the query: ("nucleic acids research"[Journal]) AND (("D1"[Issue]) OR ("database issue"[Issue])) on the 2.05 million papers in the Bridge2AI TKG, we retrieved 1,868 dataset-related papers published between 2004 and 2023. After excluding papers with missing values, we refined our dataset to include 1,699 biomedical dataset papers from NAR database issues.

### 2.3 Measurement of Scientific Impact and Clinical Translation Power

We assessed the quality and impact of each paper through two metrics: (1) Scientific Impact and (2) Clinical Translation Power. Scientific impact was measured by the Relative Citation Ratio (RCR) [14], which normalizes citation counts based on the field and publication time, benchmarking them against PubMed publications. The RCR percentile represents the scientific impact of each paper. Clinical Translation Power measures the extent to which research papers influence clinical trials and guidelines, thereby potentially contributing to successful disease treatments. This is calculated by the Approximate Potential to Translate (APT) [15], a machine learning-based estimation of a paper's likelihood to be cited in subsequent clinical trials or guidelines, with scores falling within five categories: 0.05, 0.25, 0.50, 0.75, and 0.95. An APT score of 0.05 indicates a low likelihood (5%) of being cited by clinical trials and guidelines, whereas a score of 0.95 indicates a high likelihood (95%). Papers published in NAR database issues exhibit higher scientific impact compared to the entire PubMed corpus. The median RCR percentile for these papers is 85.6, significantly higher than the PubMed median of 50. On the other hand, datasets with high clinical translation power are rare, with a mean APT of 0.32 and a median APT of 0.25. We classify papers with high citation impact as those with an RCR percentile above 74.09 (mean value). Similarly, papers with high clinical translation power are those with an APT score exceeding 0.32 (mean value).

### 2.4 Team Power and Team Diversity

The creation of high-quality biomedical datasets demands extensive teamwork, effective leadership, productivity, and diverse team composition. We developed variables to measure both team power and team fairness.

*2.4.1 Team Power.* **PI Leadership:** The Principal Investigator (PI), typically the last author, directs research and leads the team. We use the percentile of the normalized H-index of the PI to represent their leadership. **Number of Team Members:** This is the total number of researchers on the team. Larger teams benefit from more expertise but face greater challenges in communication. **Team Average H-index:** We calculate the normalized H-index percentile for each author and average these values to represent team productivity. **Team Average Career Age:** For each author, we determine the time from their first publication to the current one and average these values to represent team seniority. **Team H-index Diversity (Gini):** This measures inequality in the distribution of team members' H-index values. **Team Career Age Diversity (Gini):** This measures inequality in team seniority, with higher values indicating more hierarchical structures.

*2.4.2 Team Fairness.* **Ratio of Female Authors:** The average ratio of female authors in each team is 0.23 (Table 1), indicating underrepresentation. A higher ratio indicates a fairer composition. **Gender Diversity (Shannon):** This metric uses Shannon entropy to measure gender diversity. An all-male or all-female team scores

0, while mixed teams score higher. **Ratio of Underrepresented Races:** Authors are categorized into White, Hispanic, Black, and Asian. The ratio represents the proportion of Hispanic, Black, and Asian authors. **Race Diversity (Shannon):** This uses Shannon entropy to measure racial diversity within a team.

**Table 1: Summary of Team Attributes and Metadata for Analyzed Papers.**

| Variables | Count | Mean | Std | Min | 25% | 50% | 75% | Max |
|---|---|---|---|---|---|---|---|---|
| **Team** | | | | | | | | |
| *Team Power* | | | | | | | | |
| PI Leadership (Last Author h-index) | 1699 | 0.91 | 0.13 | 0.05 | 0.89 | 0.96 | 0.99 | 1.00 |
| Number of Team Members | 1699 | 12.75 | 11.76 | 1.00 | 6.00 | 9.00 | 15.00 | 139.00 |
| Team Average H-index | 1699 | 0.74 | 0.13 | 0.22 | 0.66 | 0.75 | 0.83 | 1.00 |
| Team Average Career Age | 1699 | 11.73 | 4.68 | 2.00 | 8.33 | 11.21 | 14.50 | 33.67 |
| Team H-index Diversity (Gini) | 1699 | 0.16 | 0.09 | 0.00 | 0.09 | 0.14 | 0.21 | 0.55 |
| Team Career Age Diversity (Gini) | 1699 | 0.42 | 0.13 | 0.00 | 0.33 | 0.42 | 0.50 | 0.83 |
| *Team Fairness* | | | | | | | | |
| Ratio of Female Authors | 1699 | 0.23 | 0.18 | 0.00 | 0.09 | 0.21 | 0.33 | 1.00 |
| Gender Diversity (Shannon) | 1699 | 0.43 | 0.19 | 0.00 | 0.36 | 0.49 | 0.57 | 0.67 |
| Ratio of Underrepresented Races | 1699 | 0.31 | 0.28 | 0.00 | 0.11 | 0.24 | 0.43 | 1.00 |
| Race Diversity (Shannon) | 1699 | 0.47 | 0.32 | 0.00 | 0.24 | 0.53 | 0.68 | 1.28 |
| **Paper** | | | | | | | | |
| Approximate Potential to Translate (APT) | 1699 | 0.32 | 0.30 | 0.05 | 0.05 | 0.25 | 0.50 | 0.95 |
| Percentile of Relative Citation Ratio (RCR) | 1699 | 74.09 | 27.67 | 0.00 | 55.95 | 85.60 | 97.70 | 100.00 |
| Year of Publication | 1699 | 2012.9 | 5.27 | 2004 | 2008 | 2013 | 2017 | 2022 |
| Citation Count | 1699 | 295.86 | 613.03 | 0.00 | 37.00 | 89.00 | 264.00 | 9578.00 |
| Relative Citation Ratio (RCR) | 1699 | 12.44 | 35.04 | 0.00 | 1.17 | 3.08 | 9.51 | 594.45 |
| Five-Year Citation Count | 1699 | 2.88 | 8.06 | 0.00 | 0.00 | 1.00 | 2.00 | 186.00 |
| Number of References | 1699 | 28.76 | 15.22 | 0.00 | 18.00 | 26.00 | 37.00 | 105.00 |

## 2.5 SHAP Explainable AI

To predict scientific impact (high-impact papers) and clinical translational power (high-translational power papers), we utilized team power, team fairness, and paper-level features. We employed XGBoost [20] to train and test our predictive models and applied SHAP (SHapley Additive exPlanations) [21] to interpret the relationships between team attributes and paper impact. For high scientific impact dataset paper predictions, our model achieved an F1 score of 0.72 and an ROC-AUC score of 0.788. For high clinical translational power predictions, the model yielded an F1 score of 0.70 and an ROC-AUC score of 0.714. It is important to note that, in the absence of a causal framework, we should interpret our results as correlations rather than causal relationships.

## 3 Results

We implemented two XGBoost models to predict high scientific impact and high clinical translation power for dataset papers. For each model, we used the team power variables, team fairness variables, and two paper-level variables: Year of Publication and Number of References. To interpret the contributing features, we utilized the SHAP framework, shown in Figure 1. We visualized the top eight most contributive features with SHAP outcomes. Features boxed by solid lines are team power features; features boxed by dashed lines are team fairness features.

The left panel (Figure 1A) presents SHAP outcomes for predicting high citation impact, while the right panel (Figure 1B) shows SHAP results for predicting high clinical translation power. In each case, bars (Figures 1 A1, B1) depict the average impact of each feature on the model's output. For instance, the Number of Team Members' average impact on predicting a high citation impact paper is 0.68, meaning that the Number of Team Members contributes 0.68 units of log-odds changes (Figure 1 A1). Scatter plots (Figures 1 A2, B2) illustrate the direction of correlation—positive or negative—between features and predicted outcomes for each instance, with each dot representing one instance. Lower feature values are in blue, and higher values are in pink. For instance, when predicting high citation impact papers (Figure 1 A2), the high Number of Team Members (pink dots) has a positive SHAP value (ranging from -0.1 to 2.8), indicating that larger team sizes positively contribute to the prediction of high citation impact papers. The most extreme instance has a SHAP value of 2.8, meaning that, for that case, team size contributes to a 2.8-unit increase in the log-odds for predicting high citation dataset papers.

### 3.1 Temporal Trends in Scientific Impact and Clinical Translation of Dataset Papers

Over time, biomedical dataset papers increasingly exhibit higher scientific impact and clinical translation power. The year of publication is a strong positive predictor for scientific impact and clinical translation power (Figures 1 A2, B2). Although our measurements for scientific impact and clinical translational power are time-invariant, there remains a notable positive correlation between publication year and the likelihood of a paper being impactful and translational. High-quality biological datasets are increasingly contributing to significant scientific discoveries and more effective translation into clinical trials and guidelines.

### 3.2 Team Power

*3.2.1 PI Leadership and Team Academic Prowess as Predictors of Dataset Success.* Strong PI leadership and high overall academic performance within dataset teams are positive predictors for producing impactful papers (Figure 1 A2) and achieving clinical translation (Figure 1 B2). For dataset teams, having a PI with a high H-index and team members with strong academic performance correlates with increased scientific impact and a higher likelihood of research being translated into clinical practice. In minor cases, high PI leadership negatively correlated with the dataset paper's success, indicating the complexity of strong leadership dynamics.

*3.2.2 Scientific Impact Positively Correlated with Team Size and Career Age. Clinical Translation Shows Inverse Patterns.* Team size is the strongest predictor of high citation impact for dataset papers (Figure 1 A1). Larger teams, benefiting from more expertise and resources, produced papers with higher citation impact compared to smaller teams. However, team size does not show a clear correlation with clinical translational success. The average career age of the team, signifying the team's seniority and collective experience, also positively correlates with high citation impact. However, it inversely correlates with the likelihood of being cited in clinical trials and guidelines. This indicates that although large, experienced teams excel in academic recognition, translating their findings into clinical practice presents challenges.

### 3.3 Team Fairness

*3.3.1 Positive Correlation Between Female Representation and Dataset Team's Success.* Despite the underrepresentation of females in our sample teams—demonstrated by an average Ratio of Female Authors of 0.23 and a 75th percentile of 0.33 in the 1,699 dataset paper teams (as shown in Table 1)—we observed a notable positive correlation between higher female representation and both scientific impact and clinical translation.

*3.3.2 Scientific Impact Positively Correlated with Race Diversity but Negatively with Ratio of Underrepresented Races.* Race diversity is a strong predictor of a dataset paper's citation impact (Figure 1 A2). Teams with greater racial diversity tend to receive more citations. When we examined the ratio of underrepresented races (Asian, Black, and Hispanic) within the teams, we found a negative correlation with citation impact. This suggests that while race-diverse teams achieve higher scientific impact, teams with a lower ratio of underrepresented races have a chance to perform better. This trend may reflect biases inherent in the current publication system. For clinical translation, a medium level of racial diversity is optimal for achieving clinical translation success (Figure 1 B2). The ratio of underrepresented races does not show a clear positive or negative correlation with clinical translation potential. It's crucial to emphasize that these findings represent correlations rather than causal relationships. Further investigation is required to fully understand the underlying mechanisms.

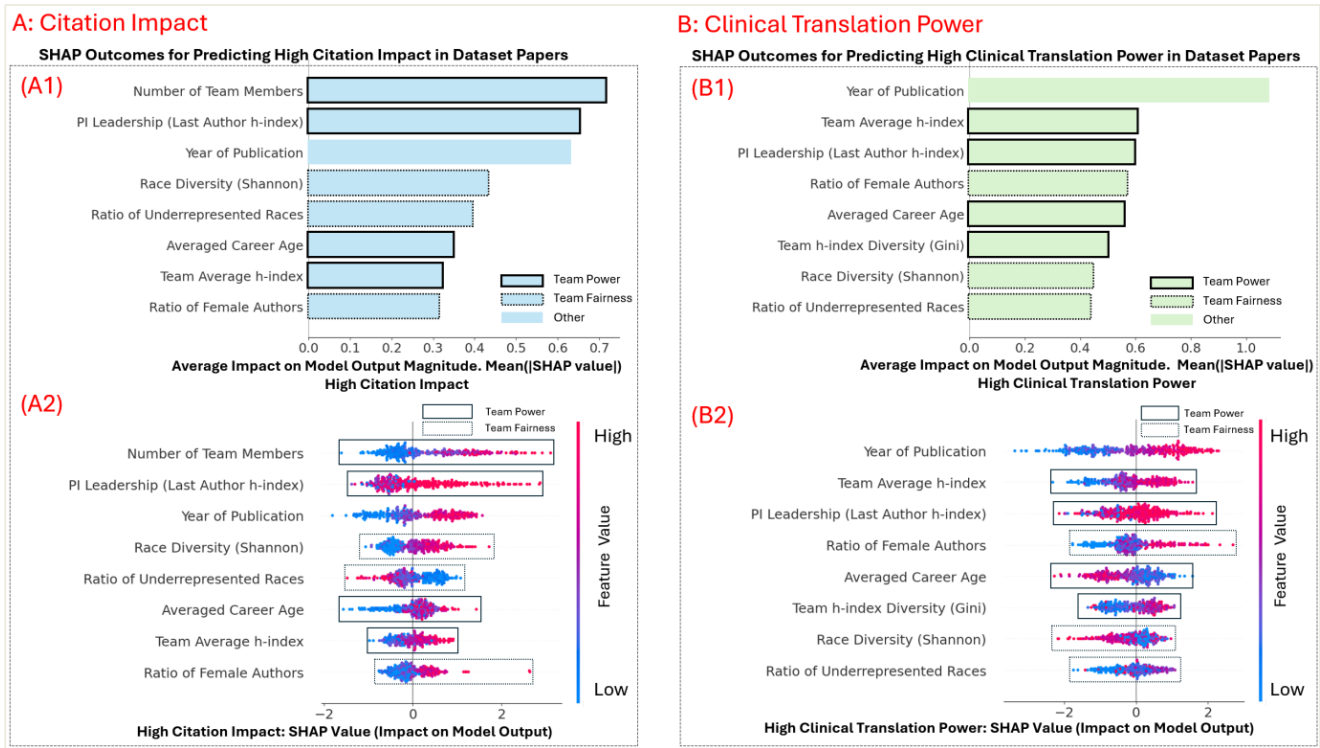

Figure 1: SHAP Outcomes for Predicting High-Impact and Clinically Translational Dataset Papers.

## 4 Conclusion

To elucidate the patterns of data generation teams that lead to clinical and scientific success, we analyzed 1,699 biomedical dataset papers from the NAR database, using the Bridge2AI Talent Knowledge Graph. We assessed the scientific impact (Relative Citation Ratio percentile) and clinical translation power (APT) of each paper, as well as team power and diversity attributes. Using XGBoost for predictions and SHAP for interpretability, our study offers crucial insights. For team power attributes, PI leadership and team academic prowess were key predictors of success. Scientific impact positively correlated with team size and career age, while clinical translation exhibited inverse patterns. For team fairness attributes, female representation correlated positively with overall success. Scientific impact was positively associated with race diversity but negatively with the ratio of underrepresented races. These findings provide strategic insights for forming teams to curate high-quality, ethical biomedical datasets. Future research should incorporate causal frameworks to delve deeper into these dynamics. Expanding the scope to include more dataset journals, such as Scientific Data, and distinguishing between original papers and updated versions of previously published datasets could further refine our understanding.

## ACKNOWLEDGMENTS

We would like to acknowledge the following funding support: NIH OT2OD032581, NIH OTA-21-008, NIH 1OT2OD032742-01, NSF 2333703, and NSF 2303038.